# Mobility of Taxol in Microtubule Bundles

Jennifer L. Ross and D. Kuchnir Fygenson

*Physics Department, University of California, Santa Barbara, CA 93106*

Mobility of taxol inside microtubules was investigated using fluorescence recovery after photobleaching (FRAP) on flow-aligned bundles. Bundles were made of microtubules with either GMPCPP or GTP at the exchangeable site on the tubulin dimer. Recovery times were sensitive to bundle thickness and packing, indicating that taxol molecules are able to move laterally through the bundle. The density of open binding sites along a microtubule was varied by controlling the concentration of taxol in solution for GMPCPP samples. With > 63% sites occupied, recovery times were independent of taxol concentration and, therefore, inversely proportional to the microscopic dissociation rate, $k_{off}$. It was found that $10\, k_{off}^{GMPCPP} \approx k_{off}^{GTP}$, consistent with, but not fully accounting for, the difference in equilibrium constants for taxol on GMPCPP and GTP microtubules. With < 63% sites occupied, recovery times decreased as $\sim [Tax]^{-1/5}$ for both types of microtubules. We conclude that the diffusion of taxol along the microtubule interior is hindered by rebinding events when open sites are within ~7 nm of each other.

**INTRODUCTION**

It has long been known (Berg, 1978) that reversible binding can alter the apparent kinetics of a chemical reaction (for a review see, (Sung et al., 1997)). Much of the interest in this subject comes from the ability of reversible binding to reduce the dimensionality of a diffusive search, and thereby enhance the rate of a biochemical reaction (Adam and Delbruck, 1968). This phenomenon, known as facilitated diffusion, is mostly studied on planar substrates (Edelstein and Agmon, 1997; Lagerholm and Thompson, 1998), for applications to cell signaling (Thompson and Axelrod, 1983) and biosensors (Ramakrishnan and Sadana, 1999). On linear substrates, the classic example is *Escherichia coli lac* repressor rapidly binding to its specific sequence on $\lambda$ DNA (Riggs et al., 1970; Berg et al., 1981; Vonhippel and Berg, 1989; Winter and Vonhippel, 1981). Facilitated diffusion has since been demonstrated and quantified for a variety of DNA binding proteins (Dowd and Lloyd, 1990; Herendeen et al., 1992; Jack et al., 1982; Nardone et al., 1986; Ricchetti et al., 1988; Surby and Reich, 1996).

Reversible binding may also lead to facilitated diffusion along the network of microtubules inside eucaryotic cells. Microtubules are hollow, cylindrical lattices of tubulin protein (Alberts et al., 1994). Each tubulin dimer has binding sites for a variety of biologically important ligands, including motor proteins (Heald and Nogales, 2002; Mandelkow and Johnson, 1998; Sosa et al., 1997; Drewes et al., 1998), microtubule associated proteins (Heald and Nogales, 2002; Drewes et al., 1998), and therapeutic drugs (Downing, 2000). The linear geometry, extraordinary stiffness and close proximity of numerous binding sites make microtubules both interesting and particularly accessible for one-dimensional transport experiments. In this paper, we assess the influence of reversible binding on the diffusion of the therapeutic drug taxol.

Taxol is common name for Paclitaxel, a microtubule-stabilizing drug important in cancer therapy (Parekh and Simpkins, 1997). Taxol binds to a site located on the inner surface of the microtubule wall (Nogales et al., 1999), where it is thought to enhance lateral contacts between dimers (Downing, 2000). Since it is small (~1 nm) compared to the inner diameter of a microtubule (~16 nm), it is expected that reversible binding, more than steric hindrance, should hinder diffusion of taxol along the microtubule axis. The effect should be especially pronounced at low taxol concentrations if the microtubule wall is impermeable (Odde, 1998). Observations of rapid association and equilibration of taxol on microtubules (Evangelio et al., 1998) originally cast doubt on the assignment of the taxol binding site to the microtubule interior. However, recent studies have shown that microtubule walls have 1-2 nm fenestrations that could allow small molecules, like taxol, to pass (Nogales et al., 1999; Meurer-Grob et al., 2001). How these pores affect mobility along the microtubule axis is an open question with implications for the biological relevance of the microtubule interior.

Here, we investigate taxol mobility using fluorescence recovery after photobleaching (FRAP) of a fluorescently labeled taxol derivative on two different kinds of microtubules: those that, like microtubules *in vivo*, have hydrolyzed the molecule of GTP at the E-site of each tubulin dimer (GTP microtubules), and those that have a non-hydrolyzable analog of GTP at the E-site (GMPCPP microtubules). These two types of microtubules are known to have equilibrium dissociation constants (Li et al., 2000) for taxol and its various derivatives that differ by two orders of magnitude, $220\, K_D^{GMPCPP} \approx K_D^{GTP}$. Our results confirm that the microtubule wall is permeable to taxol and show the fluorescence recovery times decrease with increasing taxol concentration on both types of microtubules.

## MATERIALS AND METHODS

### Tubulin and reagents

Buffers used were GTP-PEM-dex (GMPCPP-PEM-dex) (1 mM GTP (GMPCPP), 100 mM PIPES, 1 mM EGTA, 2 mM $MgSO_4$ and 4 mg/ml 110 kD dextran; buffers were pHed to 6.8 with KOH.) PIPES (free acid) was purchased from Sigma (St. Louis, MO). EGTA and $MgSO_4$ were purchased from Acros (Springfield, NJ). Dextran, molecular mass 110 kD, was purchased from Pharmacia Fine Chemicals (Uppsala, Sweden). The 4 mg/ml 110 kD dextran helped stabilize and bundle the microtubules (Rivas et al., 1999); it had a negligible effect on the viscosity of the solution. GTP (Sigma, St. Louis, MO) and GMPCPP (Guanosine-5'-[(α, β)-methyleno]triphosphate) (JenaBioScience, Jena, Germany), a non-hydrolizable analogue of GTP, were stored at 10 mM in deionized water at −20°C until needed.

Bovine brain tubulin (Cytoskeleton, Denver, CO) was kept at −70°C at a concentration of 10 mg/ml in G-PEM with 10% glycerol until within a week of use. Prior to use, 5 µl aliquots were stored in liquid nitrogen.

The stock solution of taxol (Sigma, St. Louis, MO) was 5.86 µM in DMSO (Acros, Springfield, NJ). The fluorescent taxol derivative, BODIPY 564/570 Taxol (Molecular Probes, Eugene, OR), herein referred to as botax, was made into 100 µM stock solution in DMSO and kept in the dark at −20°C until used.

**GMPCPP microtubule samples**

The binding affinity to GMPCPP microtubules, $K_D = 15 \pm 4.0$ nM (Li et al., 2000), was assumed to be the same for taxol and botax. GMPCPP microtubules were observed with six different final concentrations of taxol in solution: 25 pM, 250 pM, 2.5 nM, 25 nM, 250 nM, and 2.5 µM. The percentage of filled taxol binding sites on a microtubule at these concentrations is given by:

$$\%\text{filled} = \frac{[\text{Tax}]/K_D}{1 + [\text{Tax}]/K_D} * 100 \ , \qquad (1)$$

corresponding to roughly, 0.2%, 2%, 14%, 63%, 94%, and 99% filled binding sites on the microtubules, respectively.

To make GMPCPP microtubules, it was necessary to replace the GTP at the E-site with GMPCPP. The protocol was: 5 μl of 10 mg/ml tubulin in GTP-PEM with 10% glycerol and no taxol was incubated at 37°C for 20 minutes to polymerize microtubules. The sample was pelleted by centrifugation at 14,000 x g for 15 minutes in 37°C. The supernatant, containing GDP and GTP in solution, was removed, and the white pellet, containing GTP microtubules, was resuspended in 20 μl of GMPCPP-PEM-dex. The sample was incubated at 4°C for 15 minutes to depolymerize the GTP-microtubules and then incubated at 37°C for 20 minutes to form GMPCPP-microtubules.

To bring the sample to the desired fill ratio (Eq 1), it is necessary to fix the final concentration of taxol in solution. To do this, one must take care that the number of taxol molecules added equals the number of taxol molecules on the microtubules plus the number of taxol molecules in free solution:

$$(\text{\# taxols added}) = (\text{\# taxols on microtubule}) + (\text{\# taxols in soln}) \quad (2)$$

$$(\text{vol. taxol})[\text{Tax}]_{high} = (\text{vol. tubulin})[\text{tubulin}](\text{fill ratio}) + ((\text{vol. tubulin}) + (\text{vol. taxol}))[\text{Tax}]_{final} \quad (3)$$

Thus, by adding the appropriate volume of high concentration taxol solution, the desired fill ratio and final concentration of free taxol were achieved. For the calculation, we assumed that all the tubulin assembled into microtubules because GMPCPP microtubules have a very low background tubulin concentration (Hyman et al., 1992).

The procedure for adding botax was as follows: After polymerization of the GMPCPP microtubules, botax was added to the desired final concentration in solution, taking into account depletion due to microtubule binding (see above). The three highest taxol concentrations used a 20:80 botax:taxol solution in order to curb photodamage during the bleach (see Experimental Procedures section). The sample was allowed to equilibrate at room temperature for 20 minutes and then centrifuged at 14,000 x g for 15 minutes to pellet the microtubules. The supernatant was removed, and the pellet, which was pink due to the botax on the microtubules, was resuspended in 20 μl of the desired final concentration of botax in GMPCPP-PEM-dex. The sample equilibrated at room temperature for 20 minutes, and then was centrifuged at 14,000 x g for 15 minutes to pellet the microtubules. The supernatant was removed, and the pellet was resuspended in 10 μl of the desired final concentration of taxol in GMPCPP-PEM-dex. This procedure works well because GMPCPP microtubules are very stable and do not disassemble while manipulating them. The sample equilibrated for one hour at room temperature before being injected into the flow cell. All manipulations were performed in the dark so as not to bleach the botax fluorophore. Samples were used within four days of preparation.

**GTP microtubule samples**

The GTP microtubules used in this study are assumed to contain mostly GDP-tubulin throughout the polymer and GTP-tubulin only in a monolayer cap at the plus-end (Tran et al., 1997). Li, et al. report two equilibrium dissociation constants for taxol binding to GTP microtubules: $K_D = 61 \pm 7.0 \text{ nM}$ and $K_D = 3.3 \pm 0.54 \text{ μM}$. The higher affinity constant is associated with the GTP-tubulin and the lower affinity constant with the GDP-tubulin in the polymer. The $K_D$ for GDP microtubules is $K_D = 2.5 \pm 0.29 \text{ μM}$, which is similar to the low affinity constant in GTP microtubules (Li et al., 2000). Throughout this paper, we assume that taxol and botax bind to GTP microtubules with the same affinity, $K_D = 3.3 \pm 0.54 \text{ μM}$.

GTP microtubules were observed with four different final concentrations of taxol: 93 nM, 1.3 μM, 13 μM, and 55 μM, corresponding to 2.7%, 28%, 80%, and 94% filled binding sites on the microtubules, respectively. The range of taxol concentrations was limited by instability of GTP microtubules at low concentration and the insolubility of taxol at high concentration.

Samples of GTP microtubules were made as follows: 5 μl of 10 mg/ml tubulin in GTP-PEM with 10% glycerol and no taxol was incubated at 37°C for 20 minutes to polymerize microtubules. After polymerization, the sample equilibrated in botax for 15-20 minutes at room temperature. The first equilibration of the highest taxol concentration (55 μM) was with a 15:85 mixture of botax:taxol instead of pure botax in order to limit the DMSO to < 12%. Over 12% DMSO has been shown to form sheets instead of microtubules (Robinson and Engelborghs, 1982). The microtubules were pelletted by centrifugation at 14,000 x g for 15 minutes at 37°C. The supernatant was discarded, and the pellet, which was pink due to botax on the microtubules, was resuspended in 10 μl of GTP-PEM-dex with botax at the desired final concentration. The sample equilibrated for an hour at room temperature before being inserted into the flow cell. All manipulations were performed in the dark so as not to bleach the botax fluorophore, and samples were used within two days of preparation.

Because GTP microtubules have a relatively high critical concentration, even in the presence of taxol (~ 8 $\mu$M) (Derry et al., 1995) and undergo dynamic instability, the amount of tubulin polymerized into microtubules and the amount of taxol in solution are codependent. Upon resuspending the pellet in buffer without tubulin, microtubules will disassemble, releasing taxol into solution, to maintain the critical concentration of tubulin in solution. As a result, the final taxol concentration in solution and, consequently, the fill ratio on the microtubules increase.

In order to know the actual amount of taxol in solution, the concentration was measured *in situ*. The Sephadex beads used to bundle the microtubules (see Flow Cell Preparation), provided regions without microtubules but were porous enough to let the taxol permeate. The fluorescence intensity was measured in the center of large beads and compared to a set of standards with botax at known concentration (see Experimental Procedures for details).

**Flow cell preparation and bundle alignment**

The flow cell was made of a 25 mm x 75 mm slide, a 24 mm x 60 mm coverslip, a parafilm gasket, and G-75 Sephadex beads (particle size 40-120 µm, Sigma, St. Louis, MO). As Fig. 1 shows, a pair of 2 mm holes were drilled into the glass slide 45 mm apart. The slide and coverslip were cleaned and coated with dextran by being immersed in a 4 mg/ml solution of 110 kD dextran and blown dry. A 5 mm x 50 mm rectangle was cut from the center of a 24 mm x 60 mm piece of parafilm wax and placed on the slide so as to define a flow path between the two holes. Sephadex beads were sprinkled in the flow path to make obstacles that would catch microtubules while flowing. The coverslip was placed on top, and the flow cell was sealed by heating on a hot plate at 100°C to melt the parafilm wax.

While heating, pressure was applied to the cell to flatten the flow cell and melt the Sephadex beads. The cell was heated slide side down to preferentially melt the beads to the slide surface. This ensured that the flowing bundles would wrap around the beads at a point close to the coverslip, within the working distance of the objective.

The cell was loaded by pipetting 10 µl of sample into one of the holes. Another 10 µl of unlabeled taxol at the desired final concentration in GTP(GMPCPP)-PEM-dex was added to the cell to improve alignment and reduce background. The 2 mm holes were sealed with wax so that the sample could be observed for several days. GMPCPP microtubules were 200-800 μm long and were stable for up to a week. GTP microtubules were 200-1200 μm long, and their stability ranged from 4 hours to 2 days, depending on the taxol concentration used.

**FRAP apparatus and experimental procedures**

Experiments were performed on an inverted microscope (Olympus XI70, Tokyo, Japan) outfitted for DIC and epi fluorescence. As Fig. 2 shows, the scope was equipped with Hg arc lamps on both transmitted and epi illumination paths. Transmitted light was focused onto the sample by a water-coupled 60x objective (NA 0.9) used as a condenser. The field iris (aperture) was used to define the bleach spot. DIC was used to locate bundles and position the bleach spot. For DIC, a diffuser, a neutral density filter (ND 0.25), and a blue interference filter (488 nm) were placed in the light path so as not to bleach the sample. For bleaching, a green filter (550 nm) replaced the diffuser and the blue filter. A shutter was used to accurately time the bleach exposure, and a second shutter was used to protect the intensified CCD during bleaching to avoid damaging the phosphorous screen.

Fluorescence recovery after photobleaching was observed by epi illumination. A shutter and two neutral density filters (ND 0.1 and 0.5) were used to limit bleaching during observation. Light passed through a green excitation filter (535 nm, width 45 nm) for rhodamine and reflected off of a dichroic mirror (560 DRLP). A 60x oil-coupled objective (NA 1.4) focused the light onto the sample, exciting the BODIPY dye (peak 564 nm). The same objective collected the BODIPY emission light (peak 570nm), passed it through the dichroic mirror and a red emission filter (630 nm, width 30 nm) to an intensified CCD camera (DVC 1312 Intensicam, DVC Co., Austin, TX).

After an appropriate bundle was located by DIC, the bundle was viewed in epi fluorescence to establish a proper exposure time. DIC was used again to focus and align the bleach spot by closing the aperture and focusing the condenser. Pictures were taken at each step to document the qualities of the bundle and spot. The shutter in the epi illumination path was adjusted for the correct exposure time and synchronized with the camera's recording intervals. Images were saved directly to RAM by the computer (C-view, DVC Co., Austin, TX). Typical runs required 350 frames (~1.2 Gbytes) and lasted 30 minutes to 2 hours. Multiple runs were taken on different bundles in the same sample.

*In situ measurement of taxol concentration in GTP micrtoubule samples*

After centrifugation and resuspension, GTP microtubules will depolymerize, causing the taxol concentration in solution to increase. In order to find the actual taxol concentration in each sample, an *in situ* measurement based on the fluorescence intensity was made.

A set of standard samples with known botax concentrations were prepared for calibration purposes. Each standard is a flow cell containing Sephadex beads and no microtubules, filled with a known concentration of botax ranging from 500 pM to 5 μM. In each standard, Sephadex beads were viewed in epi-fluorescence. Pictures were taken at varying exposure times with the gain settings, magnification, and filters left unchanged. The average intensity of the bead was plotted as a function of exposure time. A linear fit was used to extract the exposure time, $t_{1/2}$, at which the average intensity of the field of view equaled half the maximum intensity, $I_{max}/2$, of the camera's dynamic range. The procedure was repeated for several beads in the same standard, and the resultant $t_{1/2}$ values were averaged. Taxol concentration was plotted as a function of average $t_{1/2}$ and found to obey: $[TAX] \propto <t_{1/2}>^{-2.0 \pm 0.4}$.

Samples of GTP microtubules were prepared with 100% botax and viewed in epi fluorescence with the same gain settings, magnification, and filters as with standards. Data was taken and analyzed on large beads in the same manner as with the standards. The taxol concentration of each sample was found using the power law the fit function for taxol concentration.

### *Photodamage*

Photobleaching of botax could result in photodamage of microtubules if the light intensity is too high or the exposure time is too long, as with other fluorescent compounds (Vigers et al., 1988). Photodamage was easily identified in DIC and fluorescence. In the damaged region, microtubules were destroyed, and fluorescence recovery did not occur. Only data runs where the bundle did not change appearance over the time of the experiment were used. If a bundle moved, shifted, disassembled, or showed signs of photodamage, the data was not used (see Results section).

**Data analysis**

Images were analyzed using NIH Image (http://rsb.info.nih.gov/nih-image/). For each bundle, a rectangular region of interest (ROI) was selected and intensities were averaged over the short dimension and plotted over the long dimension to give an intensity profile (Fig 3 A). The first image of a time series served as a baseline that was subtracted from each intensity profile to correct for permanent intensity variations along the ROI (not due to the bleach spot). The corrected profiles were fit to a gaussian curve using the equation:

$$y = a + b * e^{\left(-(x-c)^2/2d^2\right)}, \qquad (4)$$

where *a* is the vertical offset, *b* is the amplitude, *c* is the horizontal offset, and *d* is the standard deviation (Fig 3 B). All four parameters were recorded for each time step. The procedure was automated in KaleidaGraph (v. 3.51 Synergy, Reading, PA) using Apple Script (Apple Computers, Cupertino, CA).

Recovery of fluorescence was measured by fitting the decrease in amplitude over time to an exponential decay of the form:

$$y = a + b * e^{(-t/\tau)}, \qquad (5)$$

where $\tau$ is the characteristic time for decay (Fig 3 C). Data were filtered based on the percent error in the amplitude, so as to minimize the error in $\tau$.

## RESULTS

### Video-FRAP experiments

We observed the effect of unoccupied binding sites on taxol mobility in microtubules using fluorescence recovery after photobleaching (FRAP). The experimental design and data analysis used in this study were specially developed for microtubule bundles in a thick sample. Conventional FRAP experiments use a laser to bleach and observe a thin sample of isotropic media (Axelrod et al., 1976). We used video-FRAP (Kapitza et al., 1985) that takes advantage of spatial imaging to correct for bleaching of the fluorophore during observation (see Data Analysis). Previous studies have used the 2-D Fourier transform of video-FRAP images to find the diffusion constant in thick samples of isotropic media (Berk et al., 1993; Tsay and Jacobson, 1991). This technique could not be applied to microtubule samples, which were both inhomogeneous and anisotropic. Instead, we used imaging to examine the changing intensity profile along the bundle and extract a characteristic fluorescence recovery time, $\tau$, from the changing amplitude (Fig 3). It was not possible to accurately derive diffusion constants directly from the measured recovery times, due to the inhomogeneous distribution of microtubules in 3D. Nevertheless, the $\tau$ values should be inversely proportional to the diffusion constant (Kao et al., 1993).

We confirmed the robustness of our method by repeatedly bleaching a bundle (Fig 4 A). Recovery times routinely agreed for multiple bleaches at the same location, indicating that bleaching did not damage the microtubule or the taxol binding site. In cases where photodamage did occur, fluorescence never recovered (see Methods).

The accuracy of our method was tested with measurements on freely diffusing taxol in thin (2 μm) samples without microtubules. The data analysis yielded $D = 1.3 \pm 0.1 \times 10^{-6}$ cm$^2$/s, consistent with predicted values (Odde, 1998). The identical measurement in a flow cell (~100 μm thick) was ~20% faster, consistent with the results of previous 3-D FRAP experiments (Berk et al., 1993).

**Effect of bundle geometry on measured recovery times**

Within a sample, bundles varied by an order of magnitude in apparent width, depth, and brightness. These features correlate with the number of microtubules and their packing density. Brighter, thicker bundles consistently recovered slower than dimmer, thinner bundles in the same sample (Fig. 4), indicating that taxol molecules diffuse perpendicular to the bundle axis. Another result that suggests lateral diffusion in the bundle is that the widths of the gaussian intensity profiles in Figure 3 did not substantially increase with time.

**Effect of binding site density on measured recovery times**

The density of unoccupied binding sites was varied by adjusting the taxol concentration in solution. Recovery times were averaged for a variety of bundles at each taxol concentration in order to overcome variations between bundles and meaningfully compare $\tau$ values for different binding site fill-ratios.

For GMPCPP microtubules, the taxol concentrations used were from 25 pM to 2.5 µM, which resulted in 0.2% to 99% of the taxol binding sites filled (see Eq 1). Recovery times for taxol on GMPCPP bundles ranged from ~1000 to 5000 sec. As taxol concentration decreased, the average recovery time increased, indicating that mobility of taxol molecules is hindered by rebinding events on multiple unoccupied binding sites. The dependence is well approximated by a power law, $\tau \sim [\text{TAX}]^{-0.17 \pm 0.03}$. However, for taxol concentrations over 25 nM (> 63% of binding sites occupied) $\tau$ was independent of taxol concentration (Fig 5), implying that the microtubules were effectively saturated with taxol.

For GTP microtubules, taxol concentrations ranged from 93 nM to 55 µM, which resulted in 2.7% to 94% occupation of the taxol binding sites (see Eq 1). Results were essentially the same as for GMPCPP microtubules, $\tau \sim [\text{TAX}]^{-0.19 \pm 0.06}$. We suspect that the highest taxol concentration (55 $\mu$M, 94% filled sites) saturated the GTP microtubules, but could not probe higher concentrations due to the limited solubility of taxol.

**Effect of E-site nucleotide on microscopic dissociation**

GMPCPP microtubules are known to have a greater affinity for taxol than GTP microtubules (Li et al., 2000). Accordingly, recovery was always faster on GTP bundles than on GMPCPP bundles. At comparable fill ratios, including saturation, $\tau_{\text{GMPMPP}} \approx 10\, \tau_{\text{GTP}}$ (Fig 5).

When the binding sites are saturated, a bleached molecule that leaves its binding site quickly equilibrates with free solution and the vacant binding site is immediately filled by a taxol molecule from the bath. $\tau$ is then inversely proportional to the microscopic $k_{\text{off}}$ at the binding site (Lagerholm and Thompson, 1998). The observation that $\tau_{\text{GMPCPP}} \approx 10\, \tau_{\text{GTP}}$ at saturation implies that $k_{\text{off}}^{\text{GMPCPP}} \approx k_{\text{off}}^{\text{GTP}}/10$. From the $K_D$ values for GMPCPP and GTP, 15 ± 4.0 nM and 3.3 ± 0.54 $\mu$M, respectively, and the definition $K_D = k_{\text{off}}/k_{\text{on}}$, we deduce that $k_{\text{on}}^{\text{GMPCPP}} \approx (22 \pm 9.4)\, k_{\text{on}}^{\text{GTP}}$. Therefore, the taxol molecule both binds faster to GMPCPP microtubules and unbinds faster from GTP microtubules ($k_{\text{off}}^{\text{GMPCPP}} < k_{\text{off}}^{\text{GTP}}$ and $k_{\text{on}}^{\text{GMPCPP}} > k_{\text{on}}^{\text{GTP}}$).

**DISCUSSION**

The microtubule lattice is a substrate with a high density of closely spaced sites for reversible binding of specific proteins and drugs. It is thus a uniquely accessible physical system in which to study the effects of reversible binding on diffusion. We studied the mobility of taxol molecules on microtubules as a function of the density of binding sites using FRAP.

Odde has given careful consideration to the mobility of taxol at dilute concentrations down the center of a microtubule with impermeable walls. He found that the reversible binding of taxol to the microtubule should greatly reduce the diffusivity of taxol along the microtubule lumen, with steric hindrance being a secondary effect (Odde, 1998).

Recent high-resolution structural studies show that the microtubule wall is porous (Nogales et al., 1999; Meurer-Grob et al., 2001). These studies reveal pores as large as 1.5 nm x 2.5 nm in the walls of GTP-taxol microtubules and 1.5 nm x 2.0 nm in the walls of GMPCPP microtubules (Meurer-Grob et al., 2001), both of which are large enough for a taxol molecule (1 nm in diameter) to pass. When taxol traverses a pore, one expects steric hindrance to reduce the diffusivity of the particle by a factor of five (based on the diffusivity of a spherical object at the centerline of a cylinder (Renkin, 1954)). We note that a difference in pore sizes between GMPCPP and GTP microtubules cannot explain the 220-fold difference in $K_D$ (Li et al., 2000). This is because steric hindrance through pores can only change the macroscopic $k_{on}$ and $k_{off}$, but not their ratio, $K_D$.

We observe that $k_{off}$ is lower for GMPCPP microtubules than for GTP microtubules. The difference in $K_D$ values, therefore, implies that $k_{on}^{GMPCPP}$ is larger than $k_{on}^{GTP}$ (see Results). This suggests the difference in $K_D$ between GTP and GMPCPP microtubules is caused by a structural change in the taxol binding site that occurs during hydrolysis (Li et al., 2000). Our results suggest the change is one that makes the binding site both less accessible and lower affinity upon hydrolysis.

Given that the taxol binding site is located on the interior of the microtubule (Nogales et al., 1999), the results found in this study are consistent with microtubules having porous walls. The gaussian intensity profiles do not appreciably increase in width over time, indicating that there is not much macroscopic diffusion along the bundle axis. Therefore, we conclude that taxol molecules are not constrained to diffuse only along the microtubule to which they bind. Furthermore, recovery times, although reproducible for multiple runs at the same location on any one bundle, vary with bundle width. The thicker or more densely packed the bundle, the slower the recovery (Fig 4), implying taxol molecules are diffusing perpendicular to the bundle and encountering steric hindrance or perhaps binding sites on multiple microtubules.

In the absence of pores in the walls, Odde found that reversible binding reduced the diffusion of taxol by a factor of $10^6$ for very dilute taxol concentrations. On GMPCPP microtubules, at the lowest taxol concentration (25 pM), the recovery was ~$10^4$ times slower than the recovery time found in free solution. Since recovery time is inversely proportional to the diffusion coefficient, we propose that the presence of pores increases the mobility of taxol in microtubules by a factor of 100.

FRAP measurements were done on samples over a large range of taxol concentrations. At low taxol concentration (between 25 pM and 25 nM for GMPCPP microtubules and between 93 nM and 13 μM for GTP microtubules), recovery times in both GMPCPP and GTP microtubules scaled with taxol concentration, $\tau_{GMPCPP} \sim [TAX]^{-0.17 \pm 0.03}$ and $\tau_{GTP} \sim [TAX]^{-0.19 \pm 0.06}$, respectively (Fig 5). The similarity of the two power laws implies that a similar mechanism is modifying the mobility of taxol in this regime. We conclude that the power laws are evidence of hindered diffusion due to taxol molecules rebinding to the numerous open binding sites along the microtubule.

There are two main paths for taxol to find another open site: diffusing within the lumen to another site on the same microtubule or diffusing across the microtubule wall to an open binding site on a different microtubule. In order to discuss the probability of each path, we must find a way to estimate the distance to the next open binding site. A characteristic distance, λ, can be found by inverting the density of binding sites per unit volume inside a microtubule and taking the cube root,

$$\lambda = \rho^{-1/3}. \qquad (6)$$

The density, ρ, is defined as:

$$\rho = \frac{\text{number of sites per unit length}}{\text{cross-sectional area of the microtubule}} = \frac{N/l}{\pi r^2}, \quad (7)$$

with N = number of binding sites around a microtubule, $l$ = unit length of a dimer, and $r$ = the inner radius of a microtubule. For a biological microtubule, N = 13 sites, $l$ = 8 nm, and $r$ = 8.5 nm to give λ = 5.2 nm. To find the distance to the next open site at a given fill ratio, we multiply the original 13 binding sites by the percentage of open sites. For instance, a fill ratio of 90% corresponds to 10% open sites, decreasing 13 sites to 1.3 sites, to give λ = 11.2 nm.

The shortest displacement between binding sites is 3.8 nm. For fill ratios from 0.2% to 62%, λ ranges from 5.2 nm to 7.2 nm inside the microtubule. For comparison, the closest binding site on a neighboring microtubule is at least 8 nm away. The density of sites in the perpendicular direction is lower, increasing λ to 6.9 nm.

There are more binding sites along a lumenal path than perpendicular to the bundle, as given by our measure, λ. Thus, rebinding events probably occur more often along the microtubule. For a taxol molecule to rebind to a site on a neighboring microtubule, it would have to pass through two pores, each with large steric hindrance. In contrast, the lumen pathway is wide so that steric hindrance is negligible. We conclude that rebinding events on multiple microtubules are less probable than rebinding events on the same microtubule. Nonetheless, rebinding on neighboring microtubules may play a role in slowing diffusion as well, especially considering that pores are closer spaced than binding sites along the microtubule, $\lambda \sim 2.6$ nm.

At the highest taxol concentrations, we found recovery times for GMPCPP microtubules became independent of taxol concentration. The microtubule is effectively saturated with taxol when >63% of the binding sites are filled (Fig 5). (We did not determine the onset of saturation for GTP microtubules and thus restrict this discussion to GMPCPP microtubule results.)

Because the recovery times become independent of taxol concentration at saturation, the open binding sites are no longer influencing taxol mobility. Taxol molecules equilibrate with the free solution before occupying another open site, implying that the nearest neighboring sites are probably filled at saturation. From Equation 6, $\lambda > 7.2$ nm when >63% of sites are occupied, giving a measure of the distance a taxol molecule travels before escaping into solution. The distance to escape is well below the resolution limit of the microscope, $\sim 300$ nm, and could explain why the gaussian intensity profiles spread little with time.

The distance from a binding site to the next pore is < 2.6 nm away, yet the distance traveled before escape through a pore is > 7.2 nm. This suggests that the taxol molecule encounters 2-3 pores before finally passing through one. Passage through pores is most likely inhibited by the steric hindrance of the small opening. We suggest that the steric hindrance that retards taxol mobility through the pores is equivalent to non-specific binding along the microtubule. The confinement of taxol to the microtubule interior, although temporary, effectively reduces the dimensionality of the taxol's diffusive motion and consequently alters macroscopic rate constants for reactions, such as binding, in the lumen (Adam and Delbruck, 1968). This phenomenon, a kind of facilitated diffusion within microtubules, might also apply to other microtubule-based reactions.

The experimental technique introduced here is easily adapted to measure the mobility of other therapeutic drugs or microtubule associated proteins along microtubules. Such experiments could provide valuable kinetic constants for cooperative ligands, such as tau (Goode et al., 2000), or test theoretical predictions for kinesin adsorption on the microtubule surface (Lipowsky et al., 2001; Nieuwenhuizen et al., 2002; Vilfan et al., 2001). In the present work, only relative reaction rates were extracted. An extension of the technique to experiments on single microtubules rather than bundles may allow for absolute measurements.

## SUMMARY

Fluorescence recovery after photobleaching (FRAP) on taxol labeled microtubules revealed a dependence of the mobility of taxol on the density of unoccupied binding sites. Although taxol is able to move through the microtubule walls, the presence of binding sites along the microtubule is capable of hindering diffusion as long as unoccupied sites are less than ~7nm apart. This work suggests that the microtubules could increase reaction rates by reducing the diffusive search inside the microtubule lumen.

**ACKNOWLEDGEMENTS**

We benfitted from discussions with F. Brown, A. Ekani-Nkodo, J. Lew, and C. Santangelo. We are grateful to T. Mitchison for generous gifts of GMPCPP. We also thank J.J. Correia for directing us to a commercial supplier of the same. This work was supported partially by the National Science Foundation, through the CAREER program under Award No. 9985493 and through the MRL Program under Award No. DMR00-80034, and partially by an Alfred P. Sloan Foundation Fellowship (DKF).

FIGURE 1 Schematic of the flow cell. Two holes (2 mm) drilled in the slide served as inlet and outlet ports. Parafilm wax cut to have a rectangular flow path seals the coverslip to the slide when heated as described. Sephadex beads are dusted on the flow path before sealing providing obstacles to trap microtubules.

FIGURE 2 Schematic of the FRAP apparatus used. Bleaching path: the upper Hg arc lamp is used for DIC imaging and bleaching of the sample. The illumination path goes through a green filter (550 nm) and a neutral density filter (ND 0.25), off of a mirror (M), and through an aperture to reduce the size of the spot. A shutter times the bleach. A 60x water-coupled objective (NA 0.9) is used as a condenser to focus the spot onto the sample. A shutter shields the intensified CCD camera during the bleach. Observing path: the lower Hg arc follows an epi fluorescence illumination path though two neutral density filters (ND 0.1 and 0.5) and a rhodamine excitation filter (XF, 535 nm). The light is reflected off a dichroic mirror (DM) and focused onto the sample with a 60x oil-coupled objective (NA 0.9) to excite the BODIPY fluorophore (565/571 nm). The same objective collects the emitted light which passes through the dichroic mirror and through an emission filter (MF, 630 nm) to the intensified CCD. The images are captured directly to RAM by the computer.

FIGURE 3 (*A*) Time series of fluorescence recovery after photobleaching on a bundle of GTP microtubules in 5 μM taxol; example ROI shown as a dashed line. Initially, the bleach spot is very dark. After about 5 minutes, the spot is brighter and wider, and after 20 minutes, the region has recovered to its maximum intensity. (*B*) Intensity profiles of the bleach spot at two different times. At $t = 0$, intensity profile is best fit by a gaussian with negative amplitude and a width of $29 \pm 0.1$ μm. At $t = 1214$ seconds, the profile has decreased amplitude because the spot is brighter, but the width has changed little, $34 \pm 0.4$ μm. (*C*) Amplitude of the bleached spot decays exponentially with characteristic time $\tau = 329 \pm 1.9$ seconds.

FIGURE 4 Geometry of the bundle affects fluorescence recovery after photobleaching. (*A*) A GTP microtubule bundle viewed in fluorescence. The microtubule bundle begins as a single thick bundle in the upper right, and fans out into two parts: one with more microtubules (*medium*) than the other (*thin*). The thick region was bleached four times. The average recovery time was $24 \pm 2$ seconds. (*B*) Fluorescence recovery curves at the three specified locations in Fig 4 *A*. Characteristic times are 8, 15, and 21 seconds for the *thin*, *medium* and *thick* bundles, respectively.

FIGURE 5 Average recovery times, $\tau$, for GMPCPP microtubules (*filled circles*) and GDP microtubules (*open circles*) as a function of taxol concentration. Recovery times for GMPCPP microtubules are always slower than for GTP microtubules. GMPCPP microtubules in taxol concentrations from 25 pM to 25 nM exhibit a similar power law behavior as GTP microtubules in 93 nM to 55 µM taxol. For GMPCPP microtubules above 25 nM, the trend stops and the recovery times are all around 1000 seconds. The error in recovery times is the standard error due to averaging.

Figure 1

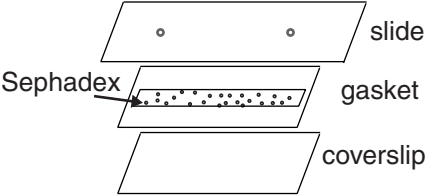

Figure 2

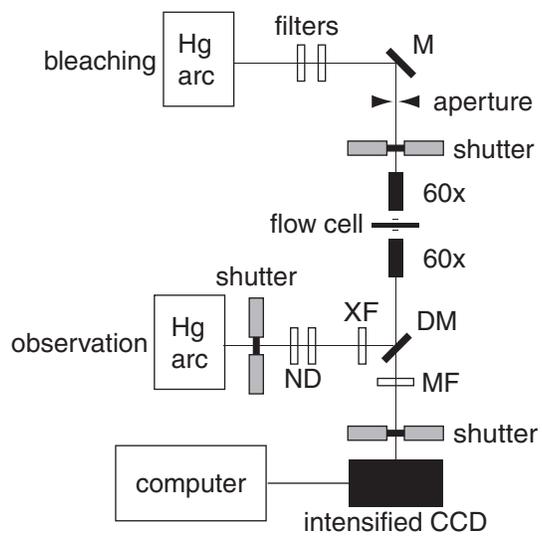

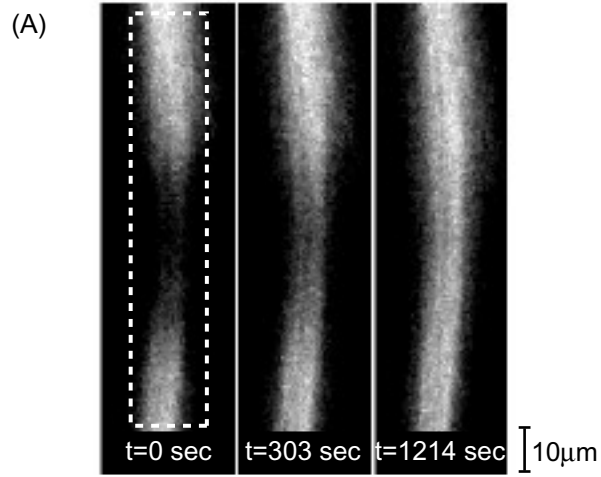

(A)

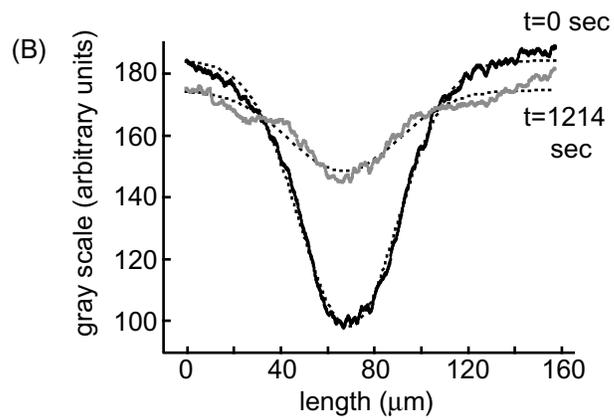

(B)

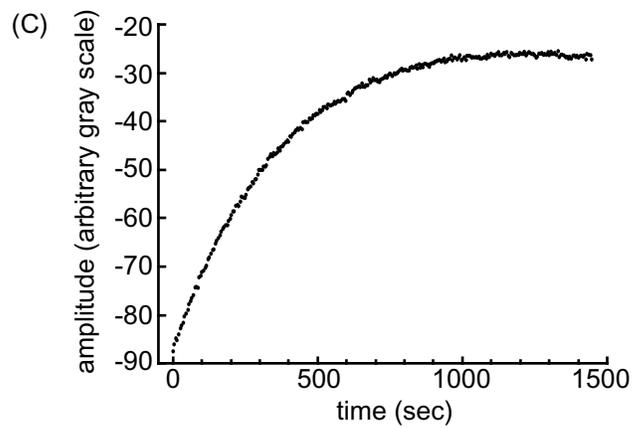

(C)

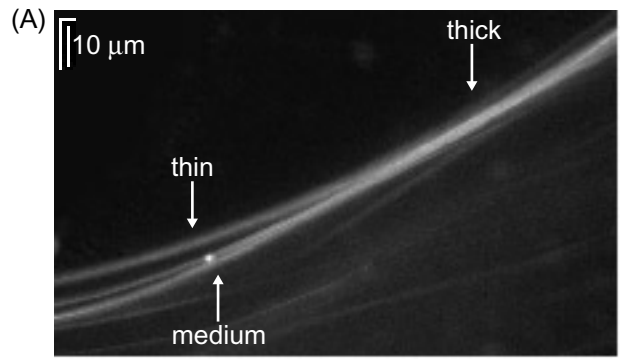
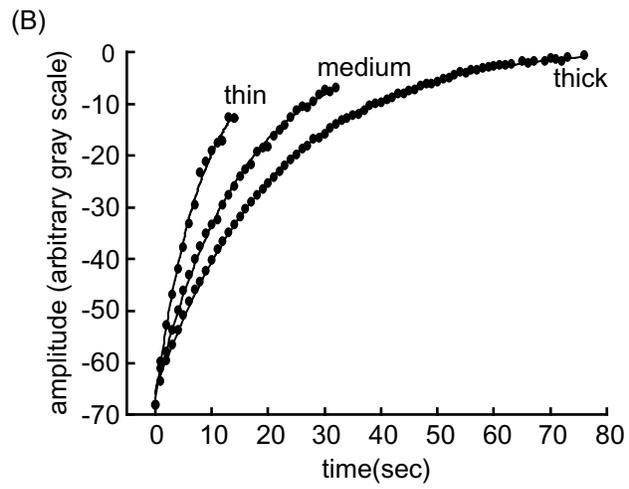

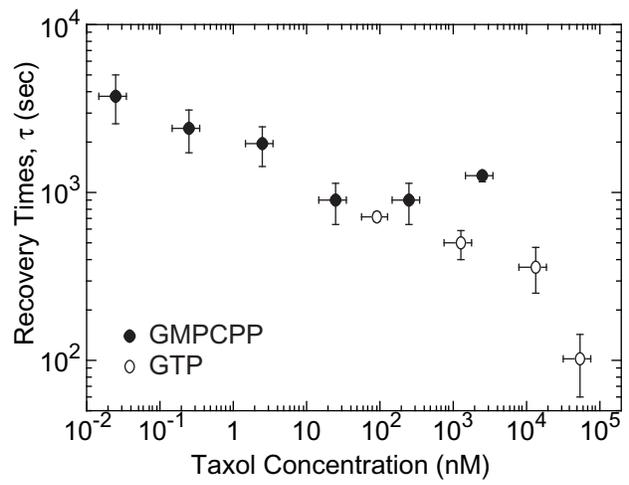